\documentclass[11pt]{article}
\usepackage{rotating}

\usepackage{lscape}

\usepackage{graphicx}
\begin{document}
\title{\bf Physical and Chemical Properties of Galactic Global Clusters
with Various Origins Identified from the Gaia DR2 Data}

\author{{V.\,A.~Marsakov,  V.\,V.~Koval', M.\,L.~Gozha}\\
{Southern Federal University, Rostov-on-Don, Russia}\\
{e-mail:  marsakov@sfedu.ru, litlevera@rambler.ru, gozha\_marina@mail.ru}}
\date{accepted \ 2020, Astronomy Reports, Vol. 64, No. 10, pp. 805-814}

\maketitle

\begin {abstract}

The differences in the relationships between the physical 
parameters and the chemical-element abundances in accreted 
globular star clusters and those formed inside the Galaxy 
have been investigated. The information on the supposed 
formation sites of the clusters based on the Gaia~DR2 data 
is borrowed from the literature. Those sources estimate 
the probability of belonging to the Galactic bulge and disk, 
as well as to six known events of the merger of dwarf satellite 
galaxies with the Milky Way, for 151~globular clusters. Orbital
elements, initial masses, population types, and ages are taken 
from the literature; the data on the chemical composition for 
69~globular clusters of the Galaxy are taken from the authors' 
compiled catalog. It is shown that all metal-poor 
($\rm{[Fe/H]}<-1.0$) genetically  related globular clusters 
have  high relative abundances of $\alpha$-elements. According 
to modern views, since type II supernovae release more 
$\alpha$-elements into the interstellar medium with increasing 
mass, it has been suggested that masses of type II supernovae 
in the Galaxy were greater than in the accreted galaxies. It 
is proved that the clusters of the low-energy group, which were 
considered accreted, are genetically related to a single 
protogalactic cloud, same as the unstratified clusters UKS~1 
and Liller~1, which most likely belong to the bulge. It is shown 
that not only the lower but also the upper limits of the clusters' 
masses decrease with an increase in the average radius of their 
orbits. The latter fact is explained by a decrease in the masses 
of emerging clusters with a decrease in the masses of their 
host galaxies. It is demonstrated that an extremely multicomponent 
stellar population is observed only in accreted globular clusters 
with an initial mass $>10^{6}M_{\odot}$. It has been suggested that these
clusters retained all the matter ejected by their evolved stars, 
from which new generations of stars formed due to long evolution 
far from our Galaxy.

\end{abstract}

{{\bf Key words:} Globular clusters, chemical composition, kinematics, 
Galaxy (Milky Way)}.

\maketitle

\section{INTRODUCTION}

According to the modern standard cosmological
model $\Lambda$CDM (Lambda-Cold Dark Matter), the
masses of galaxies grow due to mergers. The galactic
halo forms as a result of several large mergers, accompanied
by numerous small ones. When satellites merge
with a galaxy like the Milky Way, they lose their stars
due to tidal forces. These stars follow approximately
the mean orbit of their progenitor, and this leads to
stream formation.

In recent years, observational astronomy has provided
us with some compelling evidence that not all
stars that currently belong to our Galaxy formed from
the same protogalactic cloud. Some stellar objects
were captured from the nearest satellite galaxies. The
accretion epoch of extragalactic objects began at the
earliest stages of the Galaxy formation and continues
to this day \cite{1}. The presumably accreted objects can be
identified more or less reliably only after measuring
their spatial velocities and recovering their galactic
orbits. A necessary condition for verifying their extragalactic
origin is to determine their abundance of
chemical elements produced in various processes of
nuclear fusion. The fact is that isolated star-gas systems
may have different chemical evolution histories.

To identify field stars and globular clusters with a
common origin, their dynamic properties in the integral
space of motion and their chemical composition
are analyzed based on the Gaia data. Mackereth~et\,al.
\cite{2} analyzed relative abundances of $\alpha$-elements and
velocities of tens of thousands of stars within 15 kpc
from the Sun in a sample compiled by cross-identification
between the SDSS-APOGEE DR14 and Gaia
DR2 catalogs. The authors found that in metal-rich
($\rm{[Fe/H]} > -1.0$) stars with high orbital eccentricities,
the [$\alpha$/Fe] ratios are lower than those of the majority
of the nearest field stars. As a result, it was concluded
that our Galaxy at the early stages of its evolution captured
a massive ($\sim10^{9}M_{\odot}$) satellite galaxy; as a result,
part of the field stars born in the satellite entered our
Galaxy, and part of the stars of the already formed thin
disk ''heated up'', forming a subsystem of the thick disk. 
The same conclusion was reached by Helmi~et\,al.
\cite{3}, who used the APOGEE and Gaia~DR2 surveys, as
well as numerical simulations, to show that fragments
of a dwarf galaxy more massive than the Small Magellanic
Cloud, which the authors named Gaia-Enceladus,
predominate in the inner halo. They found that
among the stellar objects they studied, hundreds of
RR~Lyraes and more than a dozen globular clusters formed
in that galaxy. Moreover, the merger of the Milky Way
with Gaia-Enceladus, in their opinion, led to the
dynamic ''heating'' of the predecessor of the thick
galactic disk and, therefore, contributed to the
formation of this Galaxy subsystem approximately
10~Gyr ago. However, the Gaia-Enceladus remnants
are not the only substructure present in the neighborhood
of the Sun. In particular, the solar region is traversed
by the Helmi streams 99 discovered more than
20 years ago. Moreover, quite recently, the Gaia~DR2
and DECaPS data revealed the evidence of not one
but two captures of massive galaxies approximately 
$9-11$~Gyr ago.

Massari~et\,al. \cite{4} studied the integrals of motion
(see below for more details) and ages of Galactic globular
clusters based on the Gaia~DR2 data and the
derived parameters of galactic orbits; these studies
show that approximately 40\,\% of globular clusters
probably formed ''\textit{in situ}'', i.\,e., are genetically related
to a single protogalactic cloud. More than a third
(35\,\%) of the clusters are apparently associated with
known merger events, in particular with the Gaia-Enceladus 
galaxy (19\,\%), the Sagittarius dwarf galaxy
(5\,\%), the progenitor of the Helmi~99 streams (6\,\%),
and the Sequoia galaxy (5\,\%), although there is still
some uncertainty due to the degree of overlapping in
their dynamic characteristics. Sixteen percent of the
remaining clusters are associated with low-energy and
high-energy groups, while the rest lie in greatly elongated
orbits very high above the galactic plane, so they
are most likely of heterogeneous origin.

In \cite{5},\cite{6},\cite{7} we calculated the probabilities of the
clusters belonging to the subsystems of the thick disk
and halo using the method described by Bensby~et\,al.
\cite{8} with the aid of ground-based measurements of
their residual velocities. Only the clusters that
belonged to the thick disk were considered to be genetically
related to a single protogalactic cloud; as for the
halo, we kept only the clusters in direct orbits within
15~kpc from the galactic center. The studies showed, in
particular, that the entire set of accreted clusters in the
[Fe/H]--[$\alpha$/Fe] diagram in the range of $\rm{[Fe/H]} < -1.0$
occupies nearly the same band as fast
($V_{\Theta} < 240$~km/s), i.\,e., accreted field stars. At the same
time, almost all genetically related clusters, as well as
accreted clusters that belonged in the past to two disrupted
massive dwarf galaxies Sgt and CMa, are concentrated
towards the upper part of this band, together
with genetically related field stars ($V_{res} < 240$~km/s).

The stars of the current dwarf satellite galaxies of our
Galaxy with the same low metallicity have significantly
smaller [$\alpha$/Fe] values. As a result, it was concluded
that all stellar objects of the accreted halo are
remnants of galaxies with a higher mass than the current
environment of the Galaxy.

Since we previously identified clusters as presumably
accreted and genetically related based on error-burdened
ground-based measurements of their distances
and velocities, in this study we will use the stratification
performed by Massari~et\,al. \cite{4} based on
more accurate satellite data. These data made it possible
to determine the velocities and thus calculate the
orbits of almost all known globular clusters that currently
belong to our Galaxy, thereby significantly
increasing the number of globular clusters used for the
analysis. In this study, we are interested in the difference
between the physical and chemical properties of
sets of genetically related clusters and accreted clusters,
as well as possible differences in properties
between various presumably accreted cluster groups.


\section {INITIAL DATA}

\subsection {Chemical Element Abundances}

To analyze the behavior of some chemical elements
in globular clusters, we took spectroscopic determinations
of iron abundances and the relative abundances
of some chemical elements from our compiled catalog \cite{5},
which includes an earlier catalog by Harris \cite{9}.
Our catalog contains the averaged abundances of
28 chemical elements in the stars of 69 globular clusters
from 101 articles published from 1986 to 2018. It is
known that all the clusters underwent self-enrichment,
which changed their averaged stellar abundances
of some chemical elements (see e.\,g., \cite{10} and
references therein). The distorted abundances are
mainly of those chemical elements that are involved in
the reactions of proton captures occurring in the
hydrostatic helium burning  in the center or in
the layer sources of the asymptotic-branch giants.
When such a star sheds its shell at later evolution
stages, these elements enter the interstellar medium of
the cluster. As a result, new generations of stars have
an altered chemical composition. The average abundances
of other chemical elements in the cluster stars
remain almost unchanged (see, e.\,g., \cite{11} and references
therein). This allows us to use the averaged stellar
abundances of the undistorted chemical elements
in the clusters to study the nature of each globular
cluster.

In this paper, we consider the behavior of almost
undistorted relative abundances of four $\alpha$-elements
(magnesium, silicon, calcium, and titanium) as the
most informative in terms of diagnosing the evolution
of the early Galaxy, as well as one rapid neutron capture
element, europium. In \cite{5}, we gave a detailed description
of the procedure for averaging the relative abundances 
of each element and their errors. At the same
time, we demonstrated that the external convergence
of the abundances determined by different authors lied
in the range $\langle\sigma\rm{[el/Fe]}\rangle = (0.06-0.11)$ and deemed
possible to use our compiled abundances of chemical
elements for the statistical analysis of the initial chemical
composition of clusters belonging to different
Galactic subsystems. Since spectroscopic data are
known for less than half of globular clusters, we used
metallicities from the computer version of Harris'
compiled catalog \cite{9}, in which they are given for
almost all clusters, to analyze the relationships
between physical parameters and heavy-element
abundances.

For comparison, we used field stars from the catalog
\cite{12}, which lists the metallicities and relative abundances
of all $\alpha$-elements, as well as europium, for
785 stars of the Galaxy in the entire metallicity range
of our interest. The silicon abundances are not given;
for this reason, they are taken from the catalog \cite{13},
which contains 714 F--G field dwarfs. Unfortunately,
the latter catalog contains stars belonging mainly to
the disk populations of the Galaxy, so metal-poor stars
in it are few.

\subsection {Principles of Determining Birthplaces
of Globular Clusters}

Massari~et\,al. \cite{4} considered the positions of all
clusters in the space of calculated integrals of motion:
total orbital energies ($E$), angular momentum components
($L_{Z}$), angular momentum components ($L$) perpendicular
to, and other elements of the cluster
galactic orbits. At the same time, each cluster was analyzed
for relations to a single protogalactic cloud and
to the progenitors of the known merger events experienced
by the Galaxy. The clusters that were born, in
the authors' opinion, within the Galaxy ({\textit in situ}) were
restricted by additional parameters. The bulge clusters
should satisfy the condition $R_{max} < 3.5$~kpc (36 clusters
of this subsystem were identified as a result), while
the clusters belonging to the disk subsystem were
restricted by the low elevation of the orbital points
above the galactic plane ($Z_{max}<5$~kpc) and small
eccentricities of galactic orbits. Among the 26 identified
disk clusters, several were in retrograde orbits.
The authors found that all ''pure'' \textit{in situ} clusters show
a greater age given the same metallicity. For this reason,
they excluded two relatively young clusters (NGC~6235 and 
NGC~6254) from this group. Thus, in addition
to the integrals of motion, the ages of globular
clusters were also considered during the stratification
(see below for more details). Note that due to the large
scatter of the dynamic characteristics of the Gaia-Enceladus 
group and their overlapping with the
parameters of other groups, the authors of the cited
work had to keep double affiliation for some clusters.
As a result, in our diagrams, such clusters are indicated
by two superimposed symbols. The stratification for
all the clusters is given in [4, Table 1].

Recall that in \cite{5},\cite{6},\cite{7} we identified clusters as
genetically related (i.\,e., born from the same protogalactic
cloud) if they were in direct orbits and stayed
within 15~kpc from the galactic center. In addition,
these clusters should demonstrate the same residual
velocities as the field stars of the Galaxy's disk subsystems
(i.\,e., close to circular) and not be considered the
remnants of known satellite galaxies disrupted at that
time. Although we considered all counter-orbiting
clusters to be accreted, nevertheless, the majority of
clusters identified as genetically related using such different
approaches to kinematic data coincided (compare
Table 1 in \cite{4} and \cite{6}). Note that, unlike Massari
et\,al. \cite{4}, we added the bulge clusters to the thick disk,
while those authors had all the stellar objects of the
proper halo subsystem in the disk subsystem.

\subsection {Cluster Ages}

The ages for 68 globular clusters are also taken from
\cite{4}. The authors of that study scaled all the relative
ages determined by different authors from modern
photometric data to a uniform scale of absolute ages by
VandenBerg~et\,al. \cite{14}, which is based on the spectroscopic
scale for determining the abundances of chemical
elements by Carretta~et\,al. \cite{15}.

\subsection {Cluster Type by Multicomponent Population}

It has long been known that stellar populations in
all globular clusters are not chemically homogeneous
and contain at least two generations of stars (1G and
2G). The second population formed from the ejections
of evolved stars from the first generation, and is
characterized by increased abundances of proton capture
elements. However, there are also clusters that
show two or more parallel sequences of 1G and 2G
stars. Using the Hubble Space Telescope UV Legacy
Survey of Galactic Globular Clusters, Marino et al.
(see \cite{16} and references therein) divided the globular
clusters into two groups based on their multicomponent
stellar population. This division was made
according to spectroscopic abundances of the chemical
elements in the cluster stars. For the study, we took
all the seven clusters listed in \cite{16} with an extremely
multicomponent Type II population: NGC~362, NGC
1851, NGC 5286, NGC~6656, NGC~6715 (M~54),
NGC~7089 (M~2), and NGC~5139 ($\omega$~Cen).

\subsection {Cluster Masses} 

Until recently, the masses of clusters were determined
from their total luminosity. Now that Gaia
DR2 data have become available and it is possible to
determine not only total velocities, but also velocities
of individual stars even for the most distant clusters,
the masses are calculated from the velocity dispersions
of these stars. Based on these data, Baumgardt~et\,al.
\cite{17} figured the following: determined the dispersions
within 154 globular clusters; calculated their current
masses and galactic orbits; and recovered their initial
masses ($M_{\rm{lni}}$) considering the loss of stars by each
cluster over its lifetime as a result of interaction with
inhomogeneities of the galactic potential and dissipation
of stars. From that study, we took the current and
initial masses of the clusters, as well as the average
radii of their orbits.

\subsection {Morphological Index} 
 
The morphological index, or the horizontal branch
color $\rm{HBR}=(B-R)/(B+V+R)$, where , $B$, $V$, $R$,
are, respectively, the number of stars at the blue end of
the horizontal branch, in the instability band, and at
the red end, is taken from our catalog \cite{5}.

\section {RELATIONS BETWEEN THE PHYSICAL
CHARACTERISTICS OF CLUSTERS}

Figure 1a shows the relationship between the age
and metallicity of clusters (the latter is taken from the
catalog \cite{9}, since it is available for all clusters). The
symbols mark the clusters belonging to all nine groups
distinguished in \cite{4}. Clusters of different groups are
indicated with different symbols; large spheres denote
the clusters formed within the Galaxy (\textit{in situ}). The
diagram clearly shows two parallel dependences differing
in age. Note that the authors of \cite{14}, who were the
first to discover this structure in the diagram, did not
find an unambiguous explanation of the nature of
these two sequences, but linked their occurrence to the
difference in the loss of gas ejected by the asymptotic-branch
giants of the clusters. It is clearly seen that with
a fixed metallicity, the genetically related globular
clusters of the bulge and disk are older than the clusters,
which, based on their spatial-kinematic properties,
the authors of \cite{4} assumed to be accreted from
several disrupted satellite galaxies. The sequence
intersection of clusters with different origin in the diagram
is observed only for the oldest metal-poor clusters.
The authors of \cite{4} considered the difference in
ages to be the main signs (along with the kinematics)
of their extragalactic origin. Clusters with an extremely
multicomponent population are highlighted in the
same diagram. It can be seen that all of them are
among accreted clusters.

In Fig.~1b, the current mass of the
clusters (black dots) and the calculated initial masses
(large symbols) dependences versus the average radii of their orbits
are plotted based on the data of \cite{17}. There is a sharp
difference between the initial and current masses of
the clusters whose average radii of the orbits lie within
the solar circle. The authors of the cited work
explained the absence of clusters with initially low
mass at small galactocentric distances by their complete
disruption to date. At the same time, with an
increase in the average radius of their orbits, the lower
limit of the initial mass monotonically increases due to
the reduced destructive effect from the bulge and the
disk of the Galaxy. Note that the excess of low-mass globular
clusters at large galactocentric distances detected
from current masses has been known for a long time
\cite{18}.

Another noteworthy feature in the figure is a
decrease in the upper boundary of the initial mass of
clusters with increasing orbital radius, which becomes
sharper outside the solar orbit radius (see the vertical
dashed line). The exception is three initially massive
clusters from the disrupted Sagittarius galaxy; one of
them (M 54) was presumably the nucleus of this galaxy.
Central clusters of galaxies have a special formation
history; they lose almost no stars due to dissipation,
and can also increase their mass due to the field
stars of the parent galaxy falling on them until the disruption
of the latter. All clusters with distant orbits
($R_{sr}>R_{\odot}$) and initial masses below the average (see
the horizontal dashed line) were accreted. It turns out
that massive globular clusters are less likely to form in
dwarf galaxies. Although the Gaia-Enceladus galaxy
was still very massive, judging by the number of massive
clusters in it, it was nevertheless inferior to our
Galaxy. This assumption is consistent with the
authors' conclusion of \cite{3}, who argued that our Galaxy
captured a galaxy with a stellar mass $\sim 6\times 10^{8}M_{\odot}$
approximately 10~Gyr ago. A special situation is with
the progenitor of the low-energy group, in which the
masses of globular clusters are even greater than those
of the disk group clusters, but the orbital radii are
smaller. The authors of \cite{4} considered this group to be
the remnants of some disrupted dwarf galaxy, since
they all lie on the metallicity-age diagram in the same
band as all accreted younger clusters (see Fig.~1a).
However, as can be seen from the same figure, the
lowest-metallicity clusters, which make up more
than half of this group (7 out of 13), lie with the oldest
disk clusters. And only more metallic clusters
($\rm{[Fe/H]} >\sim -1.6$) turn out to be younger than the
bulge and disk clusters. For this reason, we believe the
origin of this group of clusters is doubtful (see below).
Figure 1b also makes it possible to suggest the origin of
three globular clusters unstratified in \cite{4}. In particular,
the small average orbital radii and high masses of the
UKS~1 and Liller~1 clusters indicate that they most
likely belong to the bulge subsystem, while the very
distant low-mass cluster (AM~4) is almost certainly
accreted.

The large triangles in Fig.~1b mark the Type II clusters,
which, according to the authors of \cite{16}, have
extremely multicomponent populations. These clusters,
as we can see, are all accreted; initially they had
masses of more than a million solar masses, and all of
them, except for two former nuclei of satellite galaxies
($\omega$~Cen and M~54), lie in orbits with radii close to the
solar radius. All these clusters are captured from 
disrupted dwarf satellite galaxies; i.\,e., in the initial period
of their formation. They were almost unaffected by the
destructive influence of the inhomogeneities of our
Galaxy's gravitational potential. As a result, thanks
also to their large mass, they were able to retain the gas
ejected by their evolved stars and form several stellar
populations within them.

The horizontal branch morphological index
(HBR)--metallicity diagram for clusters belonging to
different groups is depicted in Fig.~1c. The diagram
shows that almost all clusters (except ESO452-SC11)
with metallicity $-1.5 < \rm{[Fe/H]} < -1.0$ belonging to
the bulge and disk lie in a narrow layer above the tilted
line drawn ''by eye'', which separates the positions of
these clusters and accreted ones (see also \cite{5, 6}). Note
that all the clusters of the low-energy group also lie
above this line. This fact confirms the assumption that
they are genetically related and formed from the same
protogalactic cloud, like the bulge and disk clusters. At
the same time, as can be seen in Fig.~1c, almost all
accreted clusters, including those with an extremely
multicomponent stellar population, lie below the tilted
line. This behavior of the clusters in the diagram is not
surprising, since such a line was first drawn by Zinn
\cite{19}, separating the clusters lying inside and outside
the solar circle. Recall that in \cite{4}, the belonging of
clusters to the bulge and the disk was limited to a small
distance from the galactic center (see above). Thus,
the repeatedly stated assumption that metal-poor
clusters with anomalously reddened horizontal
branches are of extragalactic origin is confirmed (see
\cite{20}).

Figure 1d shows the mean orbital radius-metallicity
diagram for all our clusters. This diagram demonstrates
the existence of the long-known negative radial
metallicity gradient in our Galaxy. It can be seen that
the gradient is solely due to the fact that accreted clusters
mainly have $\rm{[Fe/H]}<-1.0$ (see the horizontal
dashed line), and all genetically related clusters lie
within the solar radius (see the vertical dashed line).
Five clusters that do not satisfy this condition are
noted, and three of them belong to the disrupted Sagittarius
galaxy. This galaxy evolved in isolation longer
than other disrupted galaxies and reached a higher
metallicity. It is clearly seen that among genetically
related clusters, half of the bulge clusters and a third of
the disk clusters have low metallicities, which are
characteristic of accreted clusters. This property of
genetically related clusters confirms the conclusion of
our study \cite{7} that it is incorrect to distinguish globular
clusters as belonging to the subsystem of the thick disk
or halo by metallicity $\rm{[Fe/H]}=-1.0$, as it is usually
done (see references in \cite{7}). Note that clusters of the
low-energy group have the same metallicity range as
the \textit{in-situ} clusters, although the vast majority of them
are metal-poor clusters. Nevertheless, this behavior in
the diagram is more consistent with genetically related
clusters. All clusters with an extremely multicomponent
stellar population (except for the former M\,54
nucleus) have metallicities close to $\rm{[Fe/H]}\approx-1.5$ and
average orbital radii close to the solar galactocentric
distance.

\section {RELATIONSHIPS BETWEEN ABUNDANCES
OF CERTAIN CHEMICAL ELEMENTS
AND METALLICITY IN GLOBAL CLUSTERS} 

\subsection {Alpha Elements}

Figure 2a shows the [Fe/H]--[Mg/Fe] diagrams for
globular clusters belonging to different groups and
field stars of different origin (details below). The lower
envelope for genetically related clusters is plotted for
the reference as a broken curve drawn ''by eye''. In
addition to the bulge and disk, the same envelope is
also suitable for the low-energy group, which confirms
the assumption that its clusters originate from the
interstellar gas of the same protogalactic cloud. Both
high-energy clusters of known chemistry are also close
to this line. On the other hand, a significant part of the
clusters of the accreted Gaia-Enceladus, Sequoia,
Helmi, and Sagittarius groups are located well below
this line. The metal-rich cluster UKS~1, unstratified by
the authors of \cite{4}, turned out to be among the disk and
bulge clusters. It can be seen from Fig.~2a that genetically
related clusters and field stars in the low-metallicity
range lie in the upper part of the band occupied
by all the objects under study.

Figure~2b shows the [Fe/H]--[Si/Fe] diagram for
clusters and field stars. Here, the bulge and disk clusters
in the entire metallicity range lie above the field
stars (see the lower envelope of these clusters). Recall
that the field stars in this diagram are taken from the
catalog \cite{13}, which includes only genetically related
stars of the Galaxy's disk subsystems. The vast cluster
majority of the low-energy group are also above this
line. Both clusters from the high-energy group with 
determined chemical composition here lie
close to the lower envelope of genetically related clusters.
All clusters of the Sagittarius galaxy in the entire
metallicity range turned out to be below the lower
envelope. For the rest of the accreted groups, one
lower envelope can be drawn even lower. In terms of
silicon abundance, the unstratified cluster UKS~1 is
also among the genetically related clusters.
Figure 2c depicts the [Fe/H]--[Ca/Fe] diagram.
The calcium abundances in genetically related clusters
show a loosely ordered structure, although there are
several lines of this element in the visible range and its
abundances are determined very reliably. This is
mainly expressed in the fact that clusters of all groups
occupy the entire width of the band. As a result, the
very conditional horizontal lower envelope that we
have drawn for genetically related metal-poor clusters
turned out to be the same for nearly all clusters. In particular,
clusters of low and high energy groups also
turned out to be above the drawn line. However, as can
be seen from the diagram, some clusters of the Sagittarius
group in the entire metallicity range are located
much below this line, which means that the lower
envelope of this group is lower everywhere. The cluster
UKS~1 is at the top of the diagram and has one of the
highest calcium abundances $\rm{[Ca/Fe]}=0.4$.

In Fig.~2d, which shows the [Fe/H]--[Ti/Fe] diagram,
the lower envelope can only be confidently
drawn for clusters in the range $\rm{[Fe/H]}>-1.0$. In the
less metallic range, the scatter of titanium abundances
is so great that the lower envelopes for all groups
(except low and high energy groups) should be drawn
below all field stars parallel to the abscissa at 
the value $\rm{[Ti/Fe]}\approx 0.15$. Here, the cluster UKS~1 is also located
within the band occupied by genetically related metal-rich
clusters.

\subsection {Average Abundances of $\alpha$-Elements}

According to modern views, most atoms of all
$\alpha$-elements form in the same processes of nuclear
fusion, so it is natural to expect that the relative abundances
averaged over magnesium, silicon, calcium,
and titanium will turn out to be more reliable than
averaged values for any individual element; this will
make it possible to correctly represent the differences
in the [$\alpha$/Fe] ratios in clusters of different groups. The
[Fe/H]--[Mg, Si, Ca, Ti/Fe] diagrams for globular
clusters and field stars are shown in Fig.~3a. The number 
of clusters here is slightly less than for individual
elements, but the reliability of the average values is
higher. In the figure, this is manifested in the fact that
the bands occupied by both field stars and globular
clusters are noticeably narrower, and the differences in
the cluster positions of different groups and stars of
different origin are more distinct.

For genetically related field stars and clusters, i.\,e.,
those formed from the same protogalactic cloud,
metallicity can serve as a statistical indicator of their
age, since the total abundance of heavy elements in a
closed star-gas system (which our Galaxy can be considered
in the first approximation) is steadily increasing
over time. We assume that such are the field stars
with a residual velocity $V_{res} < 240$~km/s (see \cite{21})
(they are indicated in the diagram by small dark snowflakes).
The vast majority of field stars with higher residual
velocities (marked with gray crosses) have retro-
grade rotation (see [21, Fig.~3a]). All higher-velocity
stars can be considered accreted candidates. Note that
metal-poor ($\rm{[Fe/H]}<-1.0$) genetically related field
stars are located along the upper half of the band in our
[Fe/H]--[$\alpha$/Fe] diagrams, and this confirms the conclusion
of \cite{22} that metal-poor field stars with low relative
abundances of $\alpha$-elements are accreted. Those
authors drew the dividing line at $[\alpha/Fe]\sim 0.3$ 
for stars with $\rm{[Fe/H]} < -1.0$.
We have also drawn the lower envelope ''by eye'' (as a gray
broken line) as a reference for genetically related stars;
the position of this line, as we can see, nearly coincides
with the one proposed by the authors of \cite{22}. As for
the position of metal-rich ($\rm{[Fe/H]} > -1.0$) globular clusters
in the [Fe/H]--[$\alpha$/Fe] diagrams, we described it in
detail in \cite{7}.

The dark broken ''by eye'' line in the figure shows
the lower envelope for the clusters called \textit{in situ} in \cite{4}.
Approximately in the same place, there is also the
lower envelope in the low-metallicity range drawn ''by
eye'' as a gray line for genetically related field stars.
Only the bulge cluster NGC~6293 turned out to be in
the low-metallicity range much below this line. The
silicon abundances in this cluster were the highest
among the clusters, while all other elements invariably
show low relative abundances. We can see that
almost all clusters from the low-energy group and also,
possibly, from the high-energy group (however, there
are only two of the latter with chemical composition)
also lie above the drawn line. It turns out that clusters
with certain relative abundances of $\alpha$-elements from
the low-energy group, as well as from the high-energy
group, confidently stay among genetically related
clusters. Most clusters of all other accreted groups lie
below this line. The average abundance of four alpha
elements in metal-poor genetically related clusters
$\langle\rm{[\alpha/Fe]}\rangle = 0.36 \pm 0.03$, 
while for accreted clusters it is
less outside the error limits and amounts to
$0.29 \pm 0.02$. These values coincide with those 
for the field stars: $0.34 \pm 0.02$ and 
$0.29 \pm 0.01$, respectively.

Despite the small number, it can be seen that the
clusters of the Sagittarius (Sgr) group form a rather
narrow band, which in the entire metallicity range lies
below the lower envelope for genetically related field
stars (see the gray broken line) and below the lower
envelope for genetically related globular clusters. The
only cluster UKS~1 with known relative abundances of
$\alpha$-elements, that could not be stratified by 
the authors of \cite{4}, fell into the metal-rich 
cluster region of the
disk and bulge. It is also seen that globular clusters
with an extremely multicomponent population at
rather close metallicities show a significant difference
in the relative abundances of $\alpha$-elements.

\subsection {Rapid Neutron Capture Elements}

Figure 3b shows the diagrams of the relative abundances
of a rapid neutron capture element (europium)
versus metallicity for globular clusters belonging to
different groups, as well as field stars. It is known that
rapid neutron capture elements form in the outbursts of
the least massive type II supernovae with masses $(8-10)M_{\odot}$, 
and some of these atoms form as a result
of the merger of neutron stars \cite{23}. Since $\alpha$-elements
are also ejected into the interstellar medium by SNe\,II
(albeit with masses $>10M_{\odot}$), the [Fe/H]--[$\alpha$/Fe] and
[Fe/H]--[Eu/Fe] diagrams are quite similar integrally.
However, the [Eu/Fe] ratios in the low-metallicity
range for clusters with different origin do not show a
clear difference in positions similar to those noted for
[$\alpha$/Fe]. Here, the clusters of all groups are mixed, and
only two accreted clusters fall down far outside the
error limits, $\omega$~Cen and Rup~106. However, it can be
noted that the highest [Eu/Fe] ratios in this range are
demonstrated not by genetically related, but by
accreted clusters (mainly from the Gaia-Enceladus
galaxy). At the same time, in the range of
$\rm{[Fe/H]}>-1.0$, clusters of the bulge, disk, 
and low-energy group form a sequence at the top 
of the field star band.

Note that both metal-rich clusters from the Sagittarius
galaxy, Pal~12 and Ter~7, turned out to have the
highest [Eu/Fe] ratios. Possibly, the position of the
[Eu/Fe] ratios that are opposite to relative abundances
of $\alpha$-elements is due to the rapid that interstellar matter
in the Sagittarius dwarf galaxy was most likely
enriched mainly by low-mass SNe II supernovae,
which are the main suppliers of rapid neutron capture
elements. As a result, we can see that these clusters in
this galaxy have a deficit of $\alpha$-elements and excess of
$r$-elements. However, this assumption requires additional
research.

\section {DISCUSSION} 

Let us list the main properties of globular clusters
that formed, according to the authors of \cite{4}, in situ,
i.\,e., the bulge and disk clusters. All of them are in
orbits, the average radii of which are smaller than the
solar orbital radius (see Figs.~1b,~1d). In addition to
being older than accreted clusters given the same
metallicity (see Fig.~1a), these clusters on average turn
out to have higher relative abundances of $\alpha$-elements
(see Figs.~2,~3a). As a rule, disk and bulge clusters have
predominantly large initial masses (see Fig.~1b), since
all less massive clusters near the galactic center have
already completely dissipated into separate stars. In
addition, metal-rich clusters of these groups have
extremely red horizontal branches, while for clusters
with low metallicity they are extremely blue (Fig.~1c).
The bulge clusters (and one disk cluster) with intermediate
metallicity ($-1.3 < \rm{[Fe/H]} < -0.9$ ) have horizontal
branches of intermediate color. It turns out that
the combination of all the listed properties is inherent
for clusters that formed within the same protogalactic
cloud, i.\,e., are genetically related.

It turned out that the low-energy clusters that were
classified by the authors of \cite{4} as accreted, also have
nearly all the properties listed above. Therefore, we
conclude that they are actually genetically related,
despite the fact that all of them lie in the band occupied
by accreted clusters in the metallicity-age diagram,
and some of them turned out to be younger than
genetically related clusters with the same metallicity
(see Fig.~1a). The authors of \cite{24} came to the same
conclusion based on the discovery of high relative silicon
abundances in these clusters from the SDSS-APOGEE
data. It can be assumed that the clusters of
this group formed slightly later than the others from
isolated protogalactic fragments. Of course, this
assumption needs to be verified by numerical modeling.
The metal-rich cluster UKS~1, non-stratified in [4],
in which the abundances of all $\alpha$-elements are the
same as those of the bulge and disk clusters, most
likely also belongs to the genetically related. As can be
seen from Fig.~1b, since its mass is more than a million
solar masses, and, from Fig.~1d, the average radius of
its orbit $\approx 0.7$~kpc, it belongs to the bulge. In addition,
the small average radius of the orbit and the large mass
of the Liller~1 cluster, not stratified in \cite{4}, suggests that
it, too, most likely belongs to the bulge (see also \cite{24});
however, the very distant low-mass cluster (AM~4) is
almost certainly of extragalactic origin.

The clusters of all other groups show, on average,
noticeably lower relative abundances of $\alpha$-elements
(see Figs.~2 and~3a), which implies that they were born
in galaxies of lower masses than our Galaxy. Moreover,
it is unlikely that the low values of the [$\alpha$/Fe]
ratios in accreted clusters can be explained only by the
lower rate of star formation in these disrupted dwarf
galaxies (as is usually believed), since a significant
number of young accreted clusters, as well as genetically
related ones, have high relative abundances of 
$\alpha$-elements. In addition, the relative abundances of
europium show no such trend. Most likely, the spread
is simply due to the poor mixing of the interstellar
medium in dwarf galaxies, where SNe II of different
masses flare up in different places. It seems more reasonable
to explain the observed overestimation on
average by higher masses of type II supernovae in our
very massive Galaxy. Indeed, on the one hand,
according to modern views, the yield of $\alpha$-elements
increases with increasing mass of the presupernova
(see, e.\,g., \cite{25}). On the other hand, it is known that
low-mass supernova explosions are more likely to
occur in low-mass dwarf galaxies \cite{26}. However, the
masses of the dwarf galaxies that supplied globular
clusters to our Galaxy were in fact much larger than
the masses of the dwarf satellite galaxies surrounding it
at the present time (see, in particular, \cite{5, 6, 27}). This
conclusion was made on the basis that the values of the
[$\alpha$/Fe] ratios in the stars of the surviving dwarf satellite
galaxies are substantially less than in globular clusters
and field stars that currently belong to our Galaxy
(see, e.\,g., \cite{6}). Indeed, as shown by satellite measurements
of the distances and velocities of stellar objects,
the masses of galaxies accreted 10~Gyr ago were rather
large, but less than that of our Galaxy. We have already
noted above that the mass of the dwarf galaxy Gaia-Enceladus, 
according to the authors of \cite{3}, significantly
exceeds the mass of the Small Magellanic
Cloud, just the stellar mass in which reaches
$\approx 5 \times 10^{9}M_{\odot}$. The stellar mass 
in the Sequoia galaxy is $\sim 5 \times 10^{7}M_{\odot}$, 
while the total mass is $\sim10^{10}$\cite{2}.
Modeling the tidal tail kinematics of the stars of the
Sagittarius galaxy in \cite{28} showed that, in order to
reproduce the velocity dispersion in the stream from
this galaxy, the mass of its dark halo should be
$M=6\times10^{M_{\odot}}$. The masses of other disintegrated
galaxies that formed the groups of clusters studied here
(Sequoia, High Energy, and Helmi streams~99) are
only slightly less than the masses of those listed (see
\cite{29} and references therein). As a result, it turns out
that our Galaxy accreted the most massive satellite
galaxies at the early stages of its formation, while the
least massive ones still continue to exist autonomously.

At the same time, the disrupted dwarf galaxies,
along with the already known decrease in the lower
limit of the initial masses of the clusters with distance
from the galactic center, revealed another interesting
property: they have almost no globular clusters of large
masses (see Fig.~1d). Only a few clusters are exceptions,
and two of them are the central clusters of these
former galaxies. Moreover, the upper mass limit
decreases with an increase in the average orbital radii
of the clusters. This decrease can no longer be
explained by the destruction of clusters. Numerical
modeling shows that rather massive satellite galaxies
begin to be intensively disrupted by the tidal forces of
the Galaxy only after a significant decrease in the size
of their orbits, while less massive ones are disrupted
even at distant approaches to our Galaxy \cite{30}. In this
case, the least massive clusters are lost by dwarf galaxies
in the first place, so they mostly remain in distant
orbits. Based on the foregoing, we can assume that the
lower the mass of the host galaxy, the lower the maximum
masses of the globular clusters formed in it.

Finally, a number of the most massive clusters of
disrupted dwarf galaxies exhibit extremely multicomponent
Type II stellar populations (see Fig.~1). Among
them are: the central clusters of Sequoia ($\omega$~Cen) and
Sagittarius (M~54) galaxies; 4 clusters from the most
massive disrupted galaxy Gaia-Enceladus; and the
most metal-poor cluster of this group, the disk cluster
M~22, whose horizontal branch only slightly differs
from the extreme blue. Moreover, all these clusters are
less than 12.5 Ga in age and have intermediate metallicity
($-1.7 < \rm{[Fe/H]}< -1.2$) and average radii,
approximately equal to the solar orbital radius. The
formation of extremely multicomponent stellar populations
in massive accreted clusters can be explained by
the fact that they evolved for a long time far from the
destructive influence of the inhomogeneities of the
our Galaxy's gravitational potential and, as a result,
preserved all the enriched matter released by their
evolved stars, from which new generations of stars
formed.

\section*{ACKNOWLEDGMENTS}
The authors are grateful to Davide Massari for providing
the unpublished ages of the globular clusters and Holger
Baumgardt for providing the updated initial globular cluster
masses.

\section*{FUNDING}

The research was funded by the Southern Federal University,
2020 (Ministry of Science and Higher Education of
the Russian Federation).

\renewcommand{\refname}{REFERENCES}

\newpage

\begin{figure*}
\centering
\includegraphics[angle=0,width=0.99\textwidth,clip]{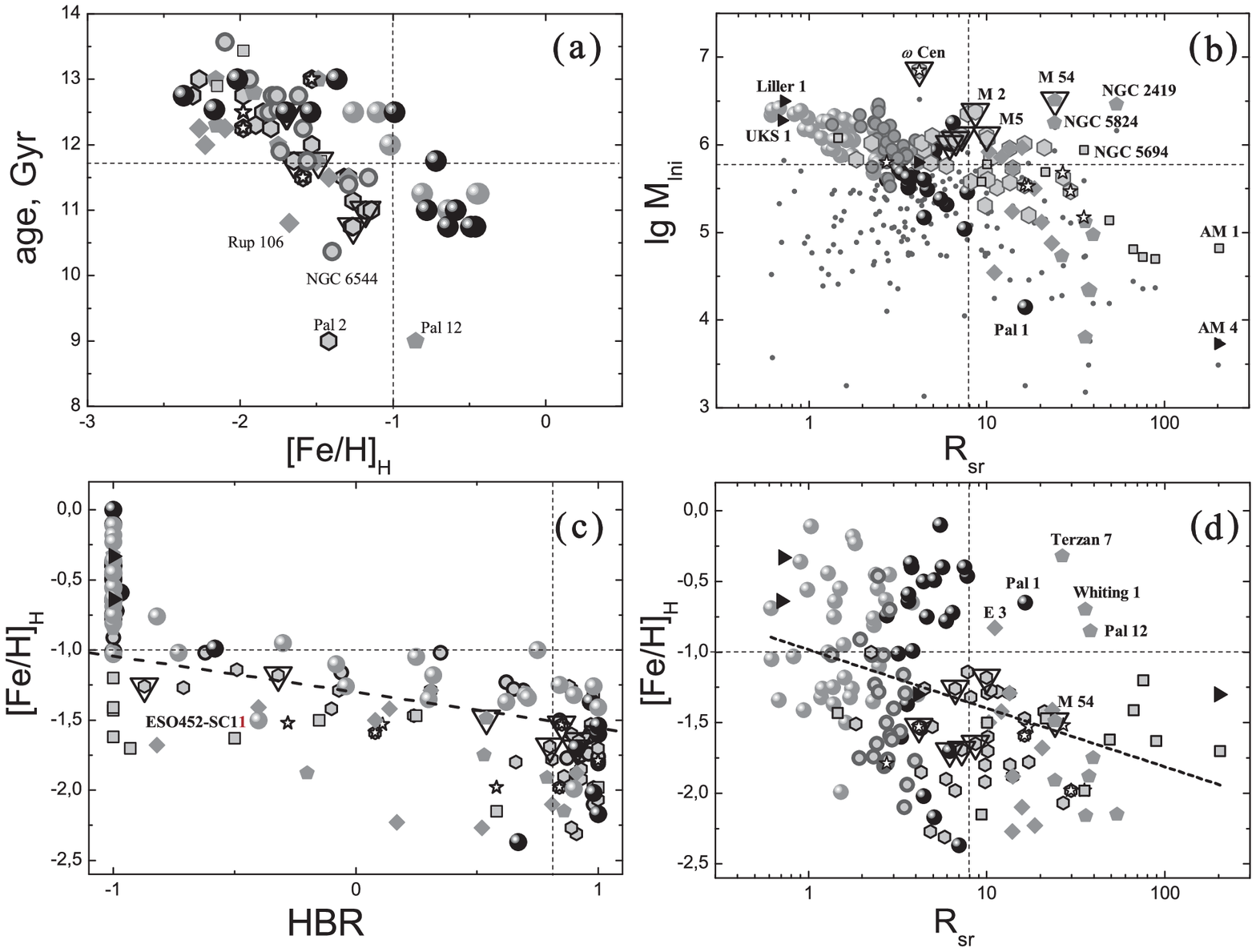}
\caption{Relationship between (a) metallicity and age; (b) 
         average orbital radius and initial mass; (c) morphological 
         index of the horizontal branch and metallicity; and (d) 
         average orbital radius and metallicity for globular 
         clusters of various origins. Genetically related globular 
         clusters are indicated by large spheres: the light spheres 
         are bulge clusters, the dark spheres are disk clusters. 
         Accreted clusters from different groups are shown by large 
         circles with light gray filling (low-energy group), 
         hexagons with light gray filling (Gaia-Enceladus), 
         gray pentagons (Sagittarius), open asterisks (Sequoia), 
         squares with light gray filling (high-energy group), 
         oblique gray squares (Helmi stream). The gray rightward 
         triangles indicate unstratified clusters, the large 
         triangles circumscribe the clusters with an extremely 
         multicomponent Type II stellar population. The gray 
         dots are the current masses of globular clusters (b). 
         The horizontal dashed lines show (a) the average age, 
         (b) average initial mass, and (c, d) traditional 
         dividing line between the disk and the halo at 
         $\rm{[Fe/H]}=-1.0$. The vertical dashed lines: the 
         average radius of the solar orbit (b, d), the line 
         HBR = 0.85 separates extreme blue clusters (c). The 
         clusters mentioned in the text are signed. The values 
         are taken from Harris' catalog \cite{9}. The inclined dashed 
         line is the radial gradient of metallicity (c, d), 
         $\rm{[Fe/H]} = (-0.42 \pm 0.08)R_{sr} - (0.98 \pm0.07)$ (d).}
\label{fig1}
\end{figure*}

\newpage

\begin{figure*}
\centering
\includegraphics[angle=0,width=0.99\textwidth,clip]{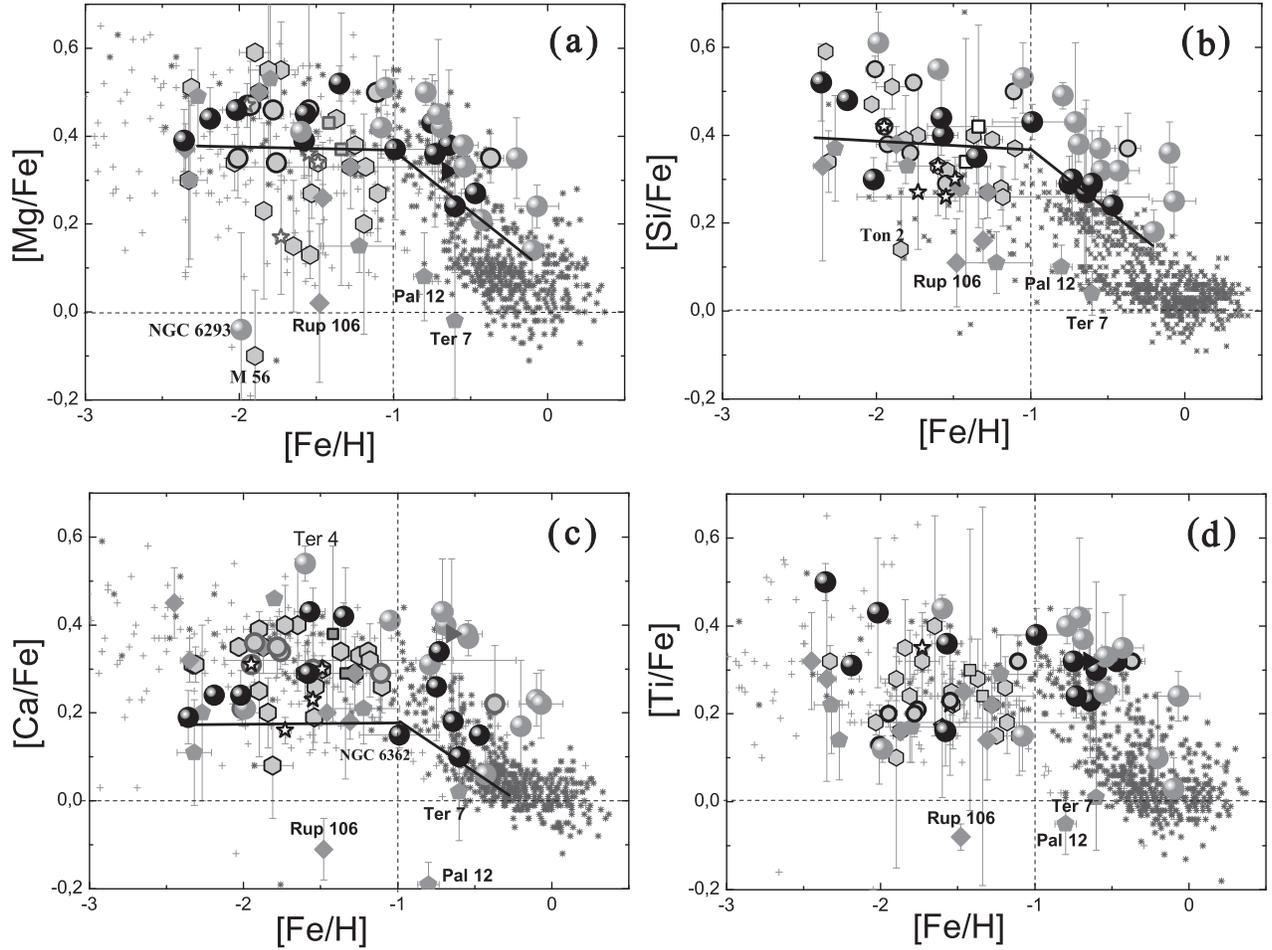}
\caption{Dependence of the relative abundances of magnesium (a), 
         silicon (b), calcium (c), and titanium (d) versus 
         metallicity for the globular clusters of our catalog 
         and field stars from the catalogs \cite{12} (a, c, d) 
         and \cite{13} (b). The notations of the globular clusters 
         and field stars are the same as in Fig.~1. The dark 
         snowflakes are genetically related field stars with 
         $V_{res} < 240$~km/s; the light gray crosses are 
         high-velocity field stars (b). The horizontal dashed 
         lines are drawn through the solar relative abundances 
         of the elements, and the vertical lines through the 
         dividing value $\rm{[Fe/H]=-1.0}$. The bars indicate 
         the authors' averaged errors in determining the 
         abundances for the globular clusters. The broken 
         curves are the lower envelopes drawn ''by eye'' 
         for genetically related clusters.}
\label{fig2}
\end{figure*}

\newpage

\begin{figure*}
\centering
\includegraphics[angle=0,width=0.99\textwidth,clip]{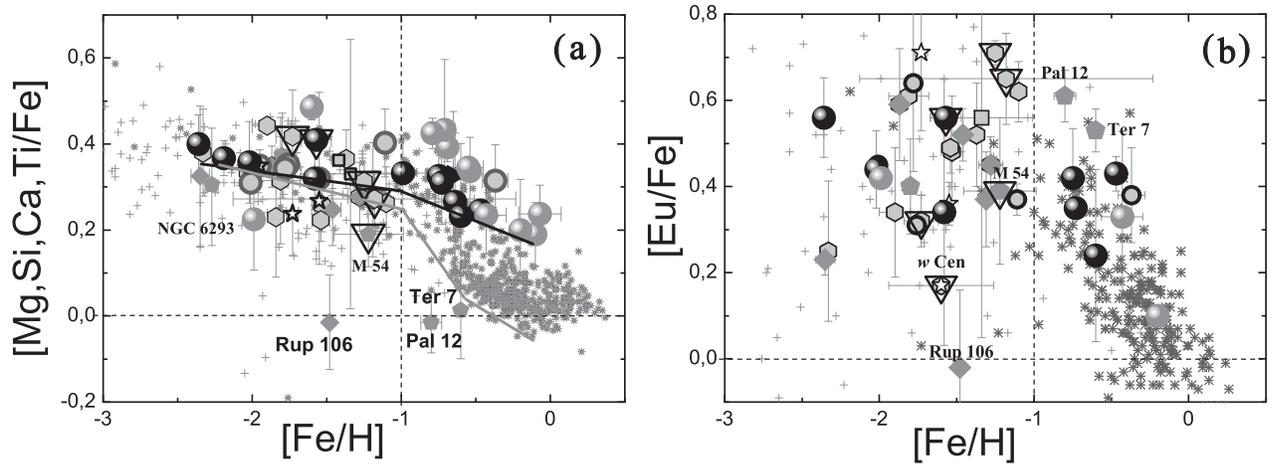}
\caption{Dependence of the relative abundances averaged over 
         four $\alpha$-elements (Mg, Si, Ca, Ti) (a), and the 
         rapid neutron capture element (Eu) (b) versus metallicity 
         for globular clusters and field stars from the catalog 
         \cite{12} (due to the absence of silicon abundances 
         for field stars, the [$\alpha$/Fe] ratios in panel 
         (a) are obtained for three elements). The notations 
         are the same as in Figs.~1 and 2. The broken lines 
         are the lower envelopes drawn ''by eye'' for genetically 
         related globular clusters (black) and field stars (gray).}
\label{fig3}
\end{figure*}



\end{document}